\documentclass[aps,prl,twocolumn,superscriptaddress,showpacs, amsmath, amssymb]{revtex4-1}	
\usepackage[dvips]{graphicx}
\usepackage{dcolumn}
\usepackage{bm}
\usepackage{color}

\begin{document}

\preprint{APS/123-QED}

\title{Narrow-line Cooling and Determination of Magic Wavelength of Cd }

\author{A. Yamaguchi}
\affiliation{Quantum Metrology Laboratory, RIKEN, Wako, Saitama 351-0198, Japan}
\affiliation{Space-Time Engineering Research Team, RIKEN, Wako, Saitama 351-0198, Japan}
\author{M. S. Safronova}
\affiliation{Department of Physics and Astronomy, University of Delaware, Newark, Delaware 19716, USA}
\affiliation{Joint Quantum Institute, NIST and the University of Maryland, College Park, Maryland 20742, USA}
\author{K. Gibble}
\affiliation{Quantum Metrology Laboratory, RIKEN, Wako, Saitama 351-0198, Japan}
\affiliation{Department of Physics, The Pennsylvania State University, University Park, Pennsylvania 16802, USA}
\author{H. Katori}
\affiliation{Quantum Metrology Laboratory, RIKEN, Wako, Saitama 351-0198, Japan}
\affiliation{Space-Time Engineering Research Team, RIKEN, Wako, Saitama 351-0198, Japan}
\affiliation{Department of Applied Physics, Graduate School of Engineering, The University of Tokyo, Bunkyo-ku, Tokyo 113-8656, Japan}

\date{\today}

\begin{abstract}
We experimentally and theoretically determine the magic wavelength of the (5$s^2$)$^{1}S_{0}$$-$(5$s$5$p$)$^{3}P_{0}$ clock transition of $^{111}$Cd to be 419.88(14)~nm and 420.1(7)~nm. To perform Lamb-Dicke spectroscopy of the clock transition, we use narrow-line laser cooling on the $^{1}S_{0}$$-$$^{3}P_{1}$ transition to cool the atoms to 6~$\mu$K and load them into an optical lattice. Cadmium is an attractive candidate for optical lattice clocks because it has a small sensitivity to blackbody radiation and its efficient narrow-line cooling mitigates higher order light shifts. We calculate the blackbody shift, including the dynamic correction, to be fractionally 2.83(8)$\times$10$^{-16}$ at 300~K, an order of magnitude smaller than that of Sr and Yb. We also report calculations of the Cd $^1P_1$ lifetime and the ground state $C_6$ coefficient.  
\end{abstract}


\maketitle

State-of-the-art optical atomic clocks deliver fractional accuracy and frequency stability of order 10$^{-18}$ \cite{Bloom-Nature-2014, Ushijima-NaturePhoton-2015, Huntemann-PRL-2016, McGrew-Nature-2018, Brewer-arXiv-2019}. Such advanced atomic clocks motivate an optical redefinition of the second \cite{Targat-NatureComm-2013} and open up new applications, such as a chronometric leveling \cite{Takano-NaturePhoton-2016, Grotti-NaturePhysics-2018} and laboratory searches for variations of fundamental constants \cite{Huntemann-PRL-2014, Safronova-RMP-2018}. At this accuracy level, one of the limiting  systematic uncertainties is the ac Stark shift of atomic clock transitions induced by black body radiation (BBR) \cite{Bloom-Nature-2014, Huntemann-PRL-2016, McGrew-Nature-2018}. While interrogating atoms in a cryogenic environment has successfully reduced the BBR shift in a Hg$^{+}$ clock \cite{Oskay-PRL-2006}, a Cs microwave clock \cite{Jefferts-PRL-2014}, and Sr and Yb optical lattice clocks \cite{Ushijima-NaturePhoton-2015, Nemitz-NaturePhoton-2016}, a number of atoms have smaller sensitivities to BBR, which can enable simpler approaches and improved accuracy. These include optical lattice clocks based on Hg, Mg, Tm, and Cd \cite{McFerran-PRL-2012, Yamanaka-PRL-2015,Kulosa-PRL-2015,Sukachev-PRA-2016, Golovisin-NatComm-2019, Kaneda-OL}, ion clocks with Al$^+$, Yb$^+$, In$^+$, and Lu$^+$ \cite{Chou-PRL-2010, Huntemann-PRL-2016, Ohtsubo-OE-2017, Arnold-NatComm-2018}, Th$^{3+}$ nuclear clock \cite{Peik-EPL-2003, Campbell-PRL-2012}, and highly charged ion clocks \cite{Dzuba-PRA-2015, Kozlov-RMP-2018}.

\begin{figure}
\includegraphics[width=1 \linewidth]{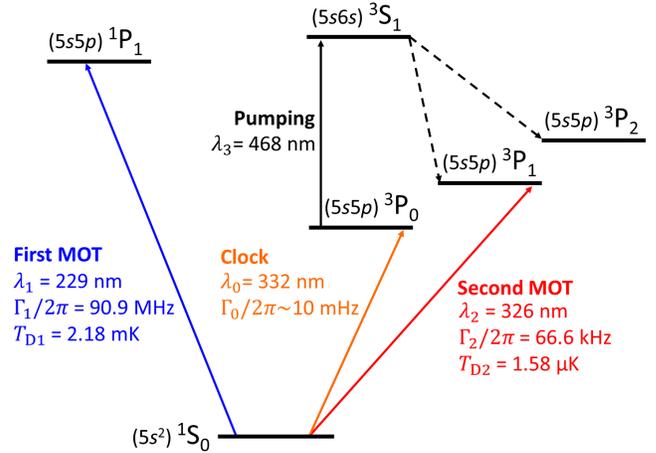}
\caption{\label{fig:EnergyLevels} Energy levels of cadmium. Wavelengths $\lambda$, natural linewidths $\Gamma/2\pi$ and, for the cooling and clock transitions, Doppler limited temperatures $T_{\rm{D}}$ are indicated. The magic wavelength for the optical lattice is 419.88 nm, at which the ac Stark shifts of both states of the $^{1}S_{0}$$-$$^{3}P_{0}$ clock transition are identical.}
\end{figure}

Among the candidates for optical lattice clocks, Cd is unique in having all of several desirable attributes. Two isotopes, $^{111}$Cd and $^{113}$Cd, both with $\gtrsim12\%$ natural abundance, have a nuclear spin of 1/2, which precludes tensor light shifts from the lattice light and provides hyperfine-induced clock transitions with natural linewidths of $\Gamma_0/2\pi = 7.0$~mHz and 7.6~mHz \cite{Garstang-JOSA-1962}. Additionally, the $\lambda_2= 326$~nm spin-forbidden $^{1}S_{0}$$-$$^{3}P_{1}$ transition's natural linewidth, $\Gamma_2/2\pi$ = 66.6~kHz \cite{Veer-Proceedings-1990, Maslowski-EPJD-2009}, allows Doppler cooling to $T_{\rm{D2}}$ = 1.58~$\mu$K, facilitating good  control of higher-order lattice light shifts \cite{Ushijima-PRL-2018}. The short wavelength of this near-ultraviolet narrow cooling transition also has applications beyond clocks; the small absorption cross section $3 \lambda_2^2/2\pi$ reduces radiation trapping \cite{Walker-PRL-1990} and allows trapping of dense cold atomic ensembles, which may enable rapid or even continuous production of quantum degenerate gases \cite{Duarte-PRA-2011, McKay-PRA-2011, Stellmer-PRL-2013}. A Cd clock can be constructed using light for all of the transitions (Fig.~\ref{fig:EnergyLevels}), including the magic wavelength, made from direct, or frequency doubled or quadrupled semiconductor lasers. Along with its insensitivity to BBR, cadmium's other favorable properties permit an optical lattice clock to be accurate, compact, and portable.

Here, we demonstrate two-stage laser cooling of Cd atoms to 6~$\mu$K using the $^{1}S_{0}$$-$$^{1}P_{1}$ transition and the spin-forbidden $^{1}S_{0}$$-$$^{3}P_{1}$ transition, shown in Fig.~\ref{fig:EnergyLevels}. We load ultracold $^{111}$Cd trapped into an optical lattice and experimentally determine the magic wavelength for the $^{1}S_{0}$$-$$^{3}P_{0}$ clock transition to be 419.88(14)~nm, in agreement with our theoretical prediction 420.1(7)~nm. Our theoretical BBR shift at 300~K is $-$0.256(7)~Hz with an extremely small dynamic correction, $-$0.45(5)~mHz. The fractional BBR shift is 2.83(8)$\times$10$^{-16}$ at 300~K, consistent with \cite{Mitroy-JPB-2010, Ovsiannikov-PRA-2016}, and allows 4$\times$10$^{-19}$ uncertainty for a temperature inaccuracy of 0.1~K.

\begin{figure}
\includegraphics[width=1 \linewidth]{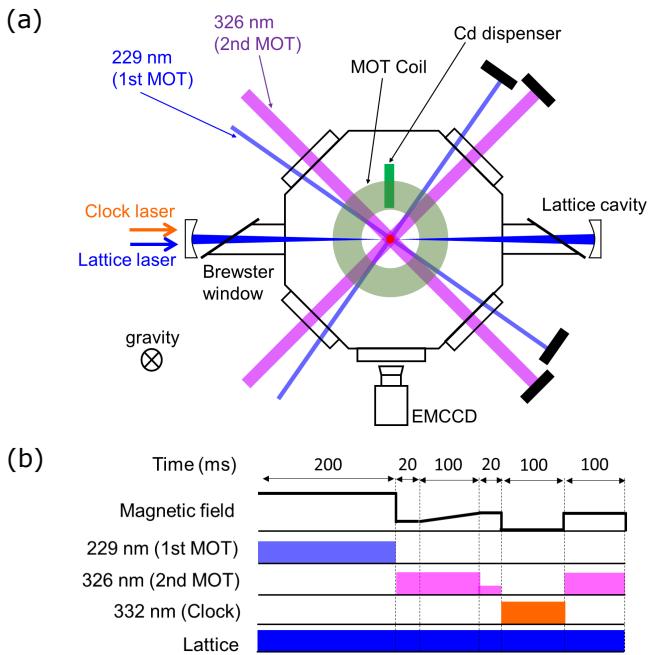}
\caption{\label{fig:Apparatus} (a) Cadmium clock schematic. The cooling and trapping lasers at 229~nm and 326~nm enter through anti-reflection-coated fused-silica viewports. The lattice light builds up in an enhancement cavity with fused-silica Brewster windows and external mirrors. (b) Sequence for laser cooling, spectroscopy and detection.}
\end{figure}

Our experimental schematic is depicted in Fig.~\ref{fig:Apparatus}(a). A $^{111}$Cd dispenser, enriched to 93\%, is located 2~cm from the magneto-optical trap (MOT). We first cool and trap $^{111}$Cd atoms using the $^{1}S_{0}$$-$$^{1}P_{1}$ transition, for which $\lambda_1$ = 229~nm and the natural linewidth is $\Gamma_1/2\pi$ = 90.9~MHz \cite{Xu-PRA-2004}. This is referred to as the first MOT \cite{Brickman-PRA-2007}, which uses a magnetic field gradient of 17~mT/cm along the axis of the anti-Helmholtz coils. It uses 6 laser beams with 1/$e^2$ radii of 1~mm and intensities of 0.2$I_{\rm{1}}$ for each, where $I_{\rm{1}}$ = 988~mW/cm$^2$ is the saturation intensity of the transition. The total laser power for the first MOT is 30~mW, which is generated by two successive second-harmonic generation (SHG) stages \cite{Kaneda-OL} fed by an external cavity diode laser (ECDL) and tapered amplifier  at 4$\lambda_1$ = 916~nm.

Figure~\ref{fig:Apparatus}(b) shows the experimental timing sequence. Operating the first MOT for 200 ms, we typically capture $10^6$ atoms. We then switch to the second MOT \cite{Katori-PRL-1999} on the narrow $\lambda_2$ = 326~nm $^{1}S_{0}$$-$$^{3}P_{1}$ transition to further cool the atoms. The laser beam 1/$e^2$ radii are 2.5~mm and the total intensity of the six beams is 310$I_{\rm{2}}$, with a saturation intensity $I_{\rm{2}}$ = 252~$\mu$W/cm$^2$. The total laser power of 50~mW is generated by SHG of a tapered amplifier seeded by an ECDL. At the beginning of the second MOT, the magnetic field gradient is reduced to 0.1~mT/cm. To capture velocities beyond the natural linewidth $\Gamma_2$, the laser frequency, tuned 3.965 MHz below the $^{1}S_{0}$$-$$^{3}P_{1}$ resonance, is sinusoidally modulated at 50~kHz with a peak-to-peak amplitude of 7.400~MHz by an acousto-optic modulator. After 20~ms, the magnetic field gradient increases to 0.3~mT/cm in 100~ms to make a compact cloud to efficiently load atoms into the optical lattice. In the last 20 ms, for optimum cooling, we inhibit the frequency modulation and set the detuning to $-3\Gamma_2$ and lower the intensity to 19$I_{\rm{2}}$.

\begin{figure}
\includegraphics[width=0.8 \linewidth]{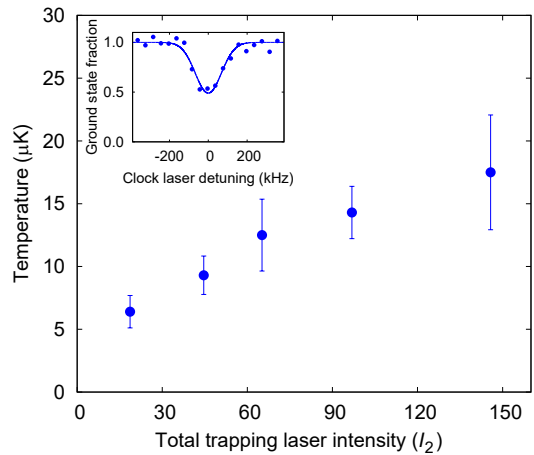}
\caption{\label{fig:AtomTemperature} Temperature of $^{111}$Cd versus laser intensity for a detuning of $-3$$\Gamma_2$. The temperature is derived from the Doppler-broadening of the $^{1}S_{0}$$-$$^{3}P_{0}$ clock transition(see inset). A Gaussian fit (solid curve) gives a minimum temperature of 6~$\mu$K for a total intensity of 19$I_{\rm{2}}$. }
\end{figure}

Figure~\ref{fig:AtomTemperature} shows the temperature of atoms from the second MOT as a function of the total laser intensity for a detuning of $-3\Gamma_2$. We measure the temperature of atoms via the Doppler broadening of the $^{1}S_{0}$$-$$^{3}P_{0}$ clock transition, as shown in the inset. The lowest temperature is 6(1)~$\mu$K for an intensity of 19$I_{\rm{2}}$. At higher intensities, the atomic temperature increases, as expected for a higher heating rate from photon scattering in Doppler cooling  \cite{Lett-JOSAB-1989}.

We load ultracold Cd into a one-dimensional optical lattice in a horizontal power enhancement cavity. The cavity finesse is 200 and the 1/$e^2$ beam radius is 71~$\mu$m. The lattice light, tunable around 420~nm, is made by SHG of light from a Ti:sapphire laser. The transfer efficiency from the second MOT to the lattice is about 15\%, loading several thousand atoms in the lattice. The $^{1}S_{0}$$-$$^{3}P_{0}$ clock transition is excited by the clock laser propagating along the lattice axis, as shown in Fig.~\ref{fig:Apparatus}(a). The 2~mW clock laser at $\lambda_0= 332$~nm is generated by SHG of an ECDL stabilized to a reference cavity. We note that the subharmonic of the Cd clock transition, $4\lambda_0= 1328$~nm, corresponds to a telecommunication wavelength, allowing dissemination of the clock signal via telecommunication fiber networks without an optical frequency comb \cite{Takano-NaturePhoton-2016}. After applying a clock laser pulse with an intensity of 25~mW/cm$^2$ for 100~ms, the excitation is observed with electron-shelving \cite{Nagourney-PRL-1986}, using the second MOT. The clock spectrum is detected as a decrease of the MOT fluorescence, which is measured with a 100~ms exposure on an electron-multiplying charge-coupled device (EMCCD) camera.

\begin{figure}
\includegraphics[width=1 \linewidth]{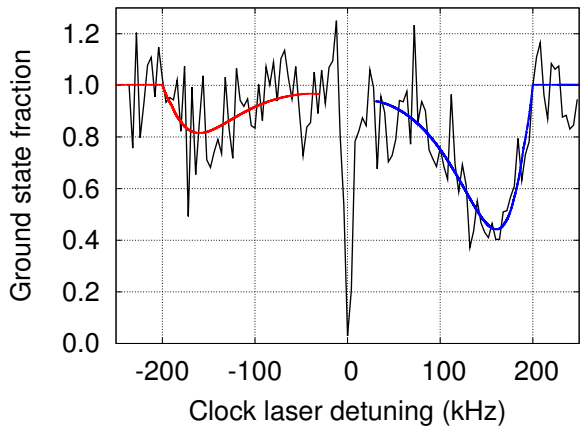}
\caption{\label{fig:LightSideband} Sideband spectrum of the clock transition for a lattice wavelength $\lambda_L=419.9$~nm. The blue and red curves are fits to the blue and red motional sidebands, which determines the axial trap frequency to be 209(8)~kHz.}
\end{figure}

To measure the lattice-trap depth, the axial trap frequency is determined from the sideband spectrum of the clock transition, as shown in Fig.~\ref{fig:LightSideband} for a lattice wavelength $\lambda_{\rm{L}}$ = 419.9~nm and peak intensity $I_{\rm{L}}\sim$ 250 kW/cm$^2$. The sideband spectra are enhanced by exciting the $^{3}P_{0}$$-$$^{3}S_{1}$ transition at $\lambda_3$ = 468 nm (see Fig.~\ref{fig:EnergyLevels}) to pump 51\% of the atoms in the $^{3}P_{0}$  state to the $^{3}P_{2}$  metastable state \cite{Takamoto-PRL-2003}. We alternately apply 2 ms pumping and clock laser pulses, to avoid light shifts and broadening, accumulating more atoms in the $^{3}P_{2}$ state and further depleting the ground state. In Fig.~\ref{fig:LightSideband}, the blue and red curves are fits to the blue and red motional sidebands  \cite{Blatt-PRA-2009}. They yield a trap frequency of 209(8)~kHz, corresponding to a lattice depth of 51(4)~$\mu$K, or 105(8)$E_{\rm{R}}$, with the lattice photon recoil energy $E_{\rm{R}} = h^2/2m\lambda_{\rm{L}}^2$ = $k_{\rm{B}}\times$0.49~$\mu$K. The asymmetry of the sideband spectra in Fig.~\ref{fig:LightSideband} indicates an atom temperature of 8(2) $\mu$K, consistent with the Doppler broadening measurement.

\begin{figure}
\includegraphics[width=1 \linewidth]{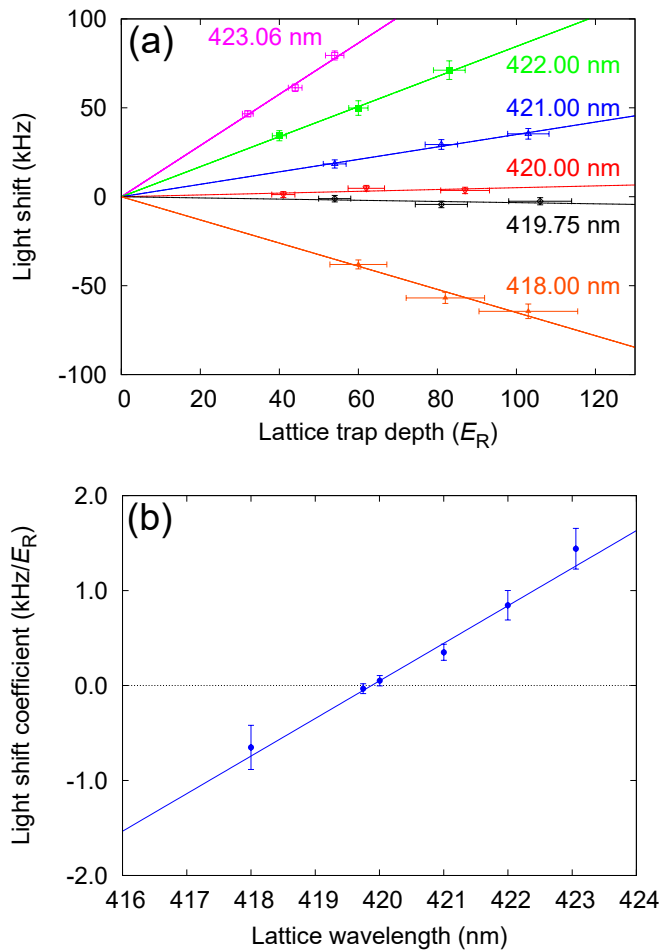}
\caption{\label{fig:LightShiftMeasurement} (a) Light shifts for six lattice wavelengths at three lattice trap depths. At each wavelength the three measured lights shifts are linearly fit to extract the light shift coefficient. The slowly drifting frequency offset is subtracted at each wavelength. (b) Light shift coefficient versus lattice wavelength. This gives a magic wavelength of 419.88(14)~nm, where the light shift coefficient goes to zero.}
\end{figure}

To determine the magic wavelength, we measure the lattice-trap-depth dependent light shift. Figure~\ref{fig:LightShiftMeasurement}(a) shows the light shift as a function of the trap depth, for six lattice wavelengths $\lambda_{\rm{L}}$, where the trap depth is extracted from sideband spectra as in Fig.~\ref{fig:LightSideband}. For clock frequency measurements, we use a single Rabi clock pulse and inhibit the 468 nm laser pulses. At each lattice wavelength, the light shifts are linearly fitted and the frequency offset of each linear fit is subtracted so that all fits intersect at zero trap depth. The typical frequency drift of the clock laser is less than 1.5~kHz in 30 minutes, the time required for light shift measurements at a single lattice wavelength. The error bars in Fig.~\ref{fig:LightShiftMeasurement}(a) include the statistical uncertainty of the fit to the center of the clock excitation spectrum and the frequency drift of the clock laser. The lattice wavelength is measured by a calibrated wavemeter with an uncertainty much less than 1~ppm.

Figure~\ref{fig:LightShiftMeasurement}(b) shows the lattice light shift per lattice-photon recoil-energy $E_{\rm{R}}$, as a function of the lattice wavelength $\lambda_{\rm{L}}$. The error bars are the 1$\sigma$ fit uncertainties from Fig.~\ref{fig:LightShiftMeasurement}(a). With a linear fit to all the light shift coefficients, the magic wavelength for the $^{1}S_{0}$$-$$^{3}P_{0}$ transition in $^{111}$Cd is $\lambda_{\rm{L}}=419.88(14)$~nm (in vacuum). The nonlinearity of the light shift is estimated to be less than 0.2\% over a wavelength range of about 10 nm. The experimentally determined magic wavelength agrees with our theoretical prediction of 420.1(7)~nm described below, and with a prior semiempirical 420~nm prediction \cite{AnpeiYe-PRA-2008}. From the linear fit of the data in Fig.~\ref{fig:LightShiftMeasurement}(b), the sensitivity of the linear ac Stark shift to the lattice laser wavelength and trap depth is 0.40(2)~kHz/$E_{\rm{R}}$/nm.

\begin{table}[t] 
\caption{\label{tab1t} Contributions to static and dynamic polarizabilities (in a.u.) at the theoretical value of the
magic wavelength, 420.1~nm. The experimental transition wavelengths $\lambda$ (in nm) and theoretical dipole matrix elements $D$ (in a.u.) are given for several leading contributions.  The last column is the differential static clock-state polarizability. }
    \begin{ruledtabular}
        \begin{tabular}{llcccc}
            \multicolumn{1}{c}{State} &
            \multicolumn{1}{c}{Contr.} &
            \multicolumn{1}{c}{$\lambda$} &
            \multicolumn{1}{c}{$D$} &
            \multicolumn{1}{c}{$\alpha_0$} &
            \multicolumn{1}{c}{$\alpha_0(\lambda_{\text{magic}})$}\\
            \hline \\[-0.8pc]
$5s^2$~$^1S_0$& $5s5p$~$^3P_1$&	326.2 & 0.158(14)&	0.12(2)	&	0.30(5)\\	
              & $5s5p$~$^1P_1$&	228.9 & 3.440(17)&	39.62(40)&	56.35(56)	\\
              & $5s6p$~$^1P_1$&	166.9  &0.689(14)&	1.16(5)	&	1.38(5)	\\
              & Other			&      &         &	0.70(5)	&	0.78(5)	\\
              &Core+vc			&      &       	&4.92(25)   & 4.92(25)\\
             &   Total			&      &      	&46.53(47)	&63.73(62) \\ [0.5pc]
$5s5p$~$^3P_0$& $5s6s$~$^3S_1$	&467.9&	1.491(11)&		15.22(23)	&	$-$63.21(95)	\\
              & $5s5d$~$^3D_1$	&340.5&	2.318(23)&		26.76(54)	&	78.0(1.6)	\\
              & $5s7s$~$^3S_1$	&308.2&	0.433(2)&		0.84(1)		&1.83(2)	     \\
              & $5s6d$~$^3D_1$	&283.8&	1.061(5)&		4.67(5)		&8.59(9)	  \\
          &     Other			&		&	    &     24.00(23)		&33.81(46)	  \\
          &   Core+vc			&		&	    &    4.70(24)		&4.70(24)	  \\
          &   Total				&		&        &   76.20(67)		&63.7(1.9) \\	[0.5pc]
  $\Delta(^3P_0-^1S_0)$  &    &         &         &29.67(82)	&
        \end{tabular}
    \end{ruledtabular}
\end{table}

Our theoretical calculation of the magic wavelength, and other Cd properties important for clock development, uses a hybrid approach that combines configuration interaction (CI) and an all-order linearized coupled-cluster method \cite{theory1}.  In this CI+all-order method, an effective Hamiltonian is constructed from a coupled-cluster calculation to account for the valence-core (vc), and core-core, correlations. This effective Hamiltonian is subsequently used in the CI calculation of the valence-valence correlations, to obtain the wave functions and the low-lying energy levels by solving the multiparticle relativistic equation $H_{\textrm{eff}}|\Psi\rangle = E|\Psi\rangle$. The valence contribution to the polarizability comes from the solution of the inhomogeneous equation from a perturbation theory of the valence space \cite{theory2}. The core polarizability and a small vc correction are calculated using the random-phase approximation. This method yielded high-precision predictions of clock-related properties of Yb \cite{theoryYb} and  Sr  \cite{theorySr}. To estimate the uncertainty of our theoretical predictions, we repeat the calculations with an effective Hamiltonian constructed from second-order perturbation theory (MBPT). The difference between this CI+MBPT and the CI+all-order gives the contributions of the dominant third and higher order terms and serves as estimates of the theoretical uncertainties. We also compare to calculations for Sr, where several matrix elements are now known with good precision, to make final estimates of our uncertainties.

The contributions to static and dynamic polarizabilities at the theoretical value of the magic wavelength are listed in Table~\ref{tab1t}. Our approach gives the total valence polarizability. Nonetheless, it is insightful to extract several of the dominant contributions, from the lowest states. The experimental transition wavelengths $\lambda$ (in nm) and dipole matrix elements $D$ (in a.u.) are given for these. The remaining contributions are grouped together as ``Other.'' While the theoretical and experimental energies agree well \cite{suppl}, it is important to use accurate experimental energies \cite{NIST} to calculate the dynamical polarizability near the magic wavelength. Because the contributions from the $5s5p$~$^3P_0 - 5s6s$~$^3S_1$ and
 $5s5p$~$^3P_0 - 5s5d$~$^3D_1$ transitions largely cancel and both resonances are close to the magic wavelength, the small experimental energy corrections can significantly shift the magic wavelength.  For consistency, we use experimental energies for the seven contributions listed in Table~\ref{tab1t}. We note that these corrections change the static polarizabilities by much less. We calculate a magic wavelength of 420.1(7)nm, where the uncertainty is the difference between the CI+all-order and CI+MBPT values.

We use the static differential clock polarizability to calculate the static contribution to the blackbody radiation shift $\Delta \nu^{\text{st}}_{\text{BBR}}=-0.255(7)$~Hz. The dynamic correction to the BBR shift is $\Delta \nu^{\text{dyn}}_{\text{BBR}}=-0.45(5)$~mHz, a factor of 330 smaller than that for Sr clocks. This gives a small fractional BBR shift $\Delta \nu_{\text{BBR}}/\nu_0$$=-2.83(8)\times 10^{-16}$. The calculations of polarizabilities and the BBR shift are described in more detail in \cite{suppl}.

We calculate the Cd $^1S_0-^1S_0$ ground state $C_6$ coefficient to be 401(8)~a.u., following \cite{theory3}. Using the theoretical $5s5p$~$^1P_1 - 5s^2$~$^1S_0$ matrix element, we predict the $^1P_1$ lifetime to be 1.500(15)~ns. We compared the calculation of $5s5p$~$^1P_1 - 5s^2$~$^1S_0$ matrix element in Cd and Sr \cite{theorySr} and find similar sign and size of the higher-order correlation effects. Due to the similarity of these two cases, confirmed by theory, we use the difference of the Sr value with the experiment ($-$0.46\%) to slightly improve the prediction of the central value of the $^1P_1$ Cd lifetime. Making the $-$0.46\% adjustment to the Cd matrix element gives  lifetimes of 3.424~a.u., or 1.514(15)~ns. The predicted $^1P_1$ lifetime is in agreement with the experimental value of  1.75(0.2)~ns \cite{Xu-PRA-2004}.

In summary, we have demonstrated two-stage laser cooling that simply and efficiently cools neutral Cd to 6~$\mu$K. Loading the ultracold Cd atoms into an optical lattice, we perform Lamb-Dicke spectroscopy on the $^{1}S_{0}$$-$$^{3}P_{0}$ clock transition and determine the magic wavelength to be 419.88(14)~nm, in agreement with our theoretical prediction of 420.1(7)~nm. We calculate the Cd blackbody shift to be 2.83(8)$\times$10$^{-16}$ at 300~K, in addition to other properties of Cd. A Cd optical lattice clock therefore can significantly improve the uncertainty originating from BBR, which currently limits the accuracy of Sr and Yb clocks. Assuming a 16-$\mu$K-deep lattice, a two-photon-ionization rate is calculated to be 2 mHz for the $^3P_0$ clock state \cite{Ovsiannikov-PRA-2016} and the Raman scattering rate to be 0.8~Hz allowing a quality factor of the clock transition of $Q\sim1.1\times10^{15}$. The convenient implementation of deep laser cooling into an optical lattice, the insensitivity to BBR, and abundant nuclear spin 1/2 isotopes make Cd an attractive candidate for compact and transportable optical clocks \cite{Koller-PRL-2017}. Further experimental investigation of higher-order polarizabilities is required to ascertain if Cd clocks can operate with lattice light shift uncertainties less than 10$^{-18}$.

We thank N. Ohmae, N. Nemitz, K. Hayashida, and M. Takamoto for their comments and technical supports. This work is supported by JST ERATO Grant No. JPMJER1002 10102832 (Japan), by JSPS Grant-in-Aid for Specially Promoted Research Grant No. JP16H06284, by JST-Mirai Program Grant Number JPMJMI18A1, Japan, by the National Science Foundation (KG), and by the USA Office of Naval Research, Grant No. N00014-17-1-2252 (MSS).

\end{document}


\title{Supplemental Material\\
Calculations of Cd polarizabilities, magic wavelength and blackbody radiation shift}

\author{A. Yamaguchi}
\affiliation{Quantum Metrology Laboratory, RIKEN, Wako, Saitama 351-0198, Japan}
\affiliation{Space-Time Engineering Research Team, RIKEN, Wako, Saitama 351-0198, Japan}
\author{M. S. Safronova}
\affiliation{Department of Physics and Astronomy, University of Delaware, Newark, Delaware 19716, USA}
\affiliation{Joint Quantum Institute, NIST and the University of Maryland, College Park, Maryland 20742, USA}
\author{K. Gibble}
\affiliation{Quantum Metrology Laboratory, RIKEN, Wako, Saitama 351-0198, Japan}
\affiliation{Department of Physics, The Pennsylvania State University, University Park, Pennsylvania 16802, USA}
\author{H. Katori}
\affiliation{Quantum Metrology Laboratory, RIKEN, Wako, Saitama 351-0198, Japan}
\affiliation{Space-Time Engineering Research Team, RIKEN, Wako, Saitama 351-0198, Japan}
\affiliation{Department of Applied Physics, Graduate School of Engineering, The University of Tokyo, Bunkyo-ku, Tokyo 113-8656, Japan}

\maketitle
\begin{table*}[b] 
\caption{\label{tab1t} Results of CI+MBPT and CI+all-order calculations. Contributions to static and dynamic polarizabilities (in a.u.) at the  theoretical value of the magic wavelength, $\lambda_{\text{magic}}=419.4$~nm for CI+MBPT and $\lambda_{\text{magic}}=420.1$~nm for CI+all-order calculations, respectively. The experimental and theoretical energies, in cm$^{-1}$, and theoretical dipole matrix elements $D$ in a.u. are given for several contributions. The difference between the experimental and theoretical energies are given in the row labeled ``Diff.,'' in cm$^{-1}$.  The dominant terms are in columns $A$ and $B$, calculated using theoretical and experimental energies respectively.}
    \begin{ruledtabular}
        \begin{tabular}{llcccccccc}
            \multicolumn{1}{c}{State} &
            \multicolumn{1}{c}{Contr.} &
            \multicolumn{2}{c}{Energy} &
                \multicolumn{1}{c}{Diff.} &
            \multicolumn{1}{c}{$D$} &
            \multicolumn{2}{c}{$\alpha_0$} &
            \multicolumn{2}{c}{$\alpha_0(\lambda_{\text{magic}})$}\\
            \hline \\[-0.8pc]
           \multicolumn{2}{c}{} &
            \multicolumn{1}{c}{Theory} &
              \multicolumn{1}{c}{Expt.} &
                \multicolumn{2}{c}{} &
            \multicolumn{1}{c}{$A$} &
            \multicolumn{1}{c}{$B$} &
              \multicolumn{1}{c}{$A$} &
            \multicolumn{1}{c}{$B$} \\
     \multicolumn{10}{c}{CI+MBPT calculations} \\
 $5s^2$~$^1S_0$ &  $5s5p$~$^3P_1$  &  31855  &  30656  &  $-$1199  &  0.172  &  0.14  &  0.14  &  0.31  &  0.36  \\
                &  $5s5p$~$^1P_1$  &  44099  &  43692  &  $-$407  &  3.426  &  38.94  &  39.30  &  55.02  &  55.97  \\
                &  $5s6p$~$^1P_1$  &  60595  &  59907  &  $-$687  &  0.675  &  1.10  &  1.11  &  1.30  &  1.32  \\
  &  Other    &    &    &    &    &  0.66  &  0.66  &  0.73  &  0.73  \\
  &  Core+vc    &    &    &    &    &  4.92  &  4.92  &  4.92  &  4.92  \\
  &  Total    &    &    &    &    &  45.76  &  46.14  &  62.29  &  63.30  \\ [0.5pc]
$5s5p$~$^3P_0$  & $5s6s$~$^3S_1$   &  20883  &  21370  &  487  &  1.502  &  15.81  &  15.45  &  $-$52.06  &  $-$63.08  \\
                & $5s5d$~$^3D_1$   &  28932  &  29372  &  440  &  2.306  &  26.89  &  26.49  &  83.82  &  77.68  \\
                & $5s7s$~$^3S_1$   &  31947  &  32449  &  502  &  0.432  &  0.86  &  0.84  &  1.93  &  1.83  \\
                & $5s6d$~$^3D_1$   &  34756  &  35239  &  483  &  1.062  &  4.74  &  4.68  &  8.96  &  8.63  \\
  &  Other    &    &    &    &    &  23.77  &  23.77  &  33.38  &  33.38  \\
  &  Core+vc    &    &    &    &    &  4.70  &  4.70  &  4.70  &  4.70  \\
  &  Total    &    &    &    &    &  76.77  &  75.93  &  80.74  &  63.15    \\[0.5pc]
   \multicolumn{10}{c}{CI+all-order calculations} \\
$5s^2$~$^1S_0$  &  $5s5p$~$^3P_1$   &  30969  &  30656  &  $-$312  &  0.158  &  0.12  &  0.12(2)  &  0.29 & 0.30(5)  \\
                &  $5s5p$~$^1P_1$   &  43715  &  43692  &  $-$22  &  3.440  &  39.60  &  39.62(40)  &  56.29 & 56.35(56)  \\
                &  $5s6p$~$^1P_1$   &  59969  &  59907  &  $-$62  &  0.689  &  1.16  &  1.16(5)  &  1.37 & 1.38(5)  \\
                &  Other            &         &         &       &         &  0.70  &  0.70(5)  &  0.78 & 0.78(5)  \\
                &  Core+vc          &         &           &     &        &  4.92   &  4.92(25)  &  4.92 & 4.92(25)  \\
               &  Total             &         &          &      &        &  46.50  &  46.53(47)  &  63.66 & 63.73(62)  \\    [0.5pc]
$5s5p$~$^3P_0$ &  $5s6s$~$^3S_1$   &  21133  &  21370  &  237  &  1.491  &  15.39  &  15.22(23)  &  $-$57.25 &  $-$63.21(95)  \\
               &  $5s5d$~$^3D_1$   &  29059  &  29372  &  313  &  2.318  &  27.05  &  26.76(54)&  82.22 & 78.0(1.6)   \\
               &  $5s7s$~$^3S_1$   &  32125  &  32449  &  324  &  0.433  &  0.85   &  0.84(1)  &  1.89 & 1.83(2)  \\
               &  $5s6d$~$^3D_1$   &  34919  &  35239  &  321  &  1.061  &  4.72   &  4.67(5)  &  8.81 & 8.59(9)  \\
               &  Other            &         &         &       &         &  24.00  &  24.00(23)  &  33.81 & 33.81(46)  \\
               &  Core+vc          &         &         &       &         &  4.70   &  4.70(24)  &  4.70 & 4.70(24)  \\
                &  Total           &          &         &      &         &  76.71  &  76.20(67)  &  74.17 & 63.7(1.9)   \\
         \end{tabular}
    \end{ruledtabular}
\end{table*}

In both our CI+all-order and CI+MBPT methods, we calculate the valence contribution to  the polarizability for a state $v$, with total angular momentum $J$ and projection $M$, from the solution of the inhomogeneous equation from a perturbation theory of the valence space \cite{theory2}
\begin{equation}
(E_v - H_{\textrm{eff}})|\Psi(v,M^{\prime})\rangle = D_{\mathrm{eff},q} |\Psi_0(v,J,M)\rangle.
\label{eq1}
\end{equation}
The effective Hamiltonian $H_{\textrm{eff}}$ includes the all-order corrections, and the effective dipole operator $D_{\textrm{eff}}$ includes random phase approximation (RPA) corrections.

The dominant contributions from the lowest states is given by the sum-over-states expression for the valence static polarizability $\alpha_0(\omega)$:
\begin{equation}
    \alpha_{0}(\omega)=\frac{2}{3(2J+1)}\sum_k\frac{\Delta E{\left\langle k\left\|D\right\|v\right\rangle}^2)}{\Delta E^2 - \omega^2}, \label{eq-1} \end{equation}
where $\Delta E = E_k-E_v$ and the sum is over intermediate states $k$ with allowed electric-dipole transitions~\cite{MitSafCla10}. These contributions are listed separately in Table~\ref{tab1t} for both the CI+MPBT and CI+all-order calculations. The remaining contributions listed in the rows ``Other'' are the difference of the total value obtained by solving the inhomogeneous equation above and the sum of the contributions calculated with the sum-over-state formula (\ref{eq-1}). The core polarizability and a small vc correction are calculated using RPA. 

We also list the theoretical and experimental energies \cite{NIST} and their differences in Table~\ref{tab1t}, along with theoretical dipole matrix elements, and contributions to the polarizability.  The results from \textit{ab initio} calculations of the dominant terms are in column ``A''.
To improve the accuracy, we recalculated these contributions using the experimental energies, which is particularly important for the dynamic polarizabilities. These are shown in column ``B'' and are used as the final values. Since the magic wavelength lies between the $5s5p$~$^3P_0 - 5s6s$~$^3S_1$ and $5s5p$~$^3P_0 - 5s5d$~$^3D_1$ resonance, these terms contribute with a opposite signs. As a result, even a 1\% change of the experimental energy leads to large changes in the total, and significantly shifts the calculated magic wavelength, for which  $\alpha_0(^1S_0)(\lambda_\text{magic})= \alpha_0(^3P_0) (\lambda_ \text{magic})$.  Using experimental energies shifts the CI+all-order magic wavelength from 425.0~nm to the final value of 420.1~nm, and for CI+MBPT, which gives less accurate energies, it shifts from 428.6~nm to a final value of 419.4~nm. We take the difference of final CI+all-order and CI+MBPT values as an uncertainty, yielding $\lambda_\text{magic}=420.1(7)$~nm.

The blackbody radiation shift (BBR) of state $v$ can be expressed in terms of the static polarizability $\alpha_0(0)$ as \cite{bbr-porsev}
\begin{equation}
\Delta E_v=-\frac{2}{15}(\alpha\pi)^3(k_BT)^4\alpha_0(0)(1+\eta),
\end{equation}
where the first term describes the static contribution and the second gives the dynamic contribution to the BBR shift. 
The quantity $\eta$ is approximated by \cite{bbr-porsev}
\begin{multline}
\eta=\eta_1+\eta_2+\eta_3=\frac{80}{63(2J+1)}\frac{\pi^2}{\alpha_0(0)k_BT}\\
\times\sum_k\frac{\vert\langle k\|D\|v\rangle\vert^2}{y^3_k}\left(1+\frac{21\pi^2}{5y_k^2}+\frac{336\pi^4}{11y_k^4}\right),
\end{multline}
where $y_k=(E_k-E_v)/(k_BT)$.

Table~\ref{bbr} lists the contributions to $\eta$ for the Cd clock states.  The contributions of $\eta_2$ and $\eta_3$ are currently negligible. We estimate the contribution of the higher states to $\eta(^3P_0)$, listed as ``Other,'' using the scaling of the higher state contributions to the $^3P_0$ polarizability. We use the final values of the static polarizabilities $\alpha_0(^1S_0)=46.53(47)$~a.u. and $\alpha_0(^3P_0)=76.20(67)$~a.u., given in Table~\ref{tab1t}, to calculate the static contribution to the BBR shift, $\Delta \nu^{\text{st}}_{\text{BBR}}=-0.255(7)$~Hz. The dynamic correction is
$\Delta \nu^{\text{dyn}}_{\text{BBR}}=-0.45(5)$~mHz, giving a total BBR shift of $-0.256(7)$~Hz, fractionally $\Delta \nu^{\text{BBR}}/\nu_0=-2.83(8)\times 10^{-16}$ at 300K.

\begin{table}[th] 
\caption{\label{bbr} Contributions to the dynamic correction $\eta$ of the BBR shift at 300K.}
    \begin{ruledtabular}
        \begin{tabular}{lll}
            \multicolumn{1}{c}{State} &
            \multicolumn{1}{c}{Contr.} &
            \multicolumn{1}{c}{$\eta$}  \\

            \hline
$5s^2$~$^1S_0$  &  $5s5p$~$^3P_1$   &0.000\,002        \\
                &  $5s5p$~$^1P_1$   &0.000\,365(3)    \\
                &  $5s6p$~$^1P_1$   &0.000\,006        \\                    
                &  Total            & 0.000\,373(3)   \\ [0.5pc]             
$5s5p$~$^3P_0$ &  $5s6s$~$^3S_1$   & 0.000\,360(5)    \\                       
               &  $5s5d$~$^3D_1$   & 0.000\,334(6)     \\                    
               &  $5s7s$~$^3S_1$   & 0.000\,009      \\
               &  $5s6d$~$^3D_1$   & 0.000\,040          \\
               &  $5s8s$~$^3S_1$   & 0.000\,002           \\
               &  $5s7d$~$^3D_1$   & 0.000\,013            \\
               &  $5s9s$~$^3S_1$   & 0.000\,001        \\
               &  $5s8d$~$^3D_1$   & 0.000\,010        \\
               &   Other           & 0.000\,148(74)      \\
               & Total             &  0.000\,918(74)   \\             
  \end{tabular}
     \end{ruledtabular}
\end{table}